# Reaping the Rewards Later: How Education Improves Old-Age Cognition in South Africa[★]

Plamen Nikolov [acde]     Steve Yeh [b★]

**Abstract**. Cognition, a component of human capital, is fundamental for decision-making, and understanding the causes of human capital depreciation in old age is especially important in aging societies. Using various proxy measures of cognitive performance from a longitudinal survey in South Africa, we study how education affects cognition in late adulthood. We show that an extra year of schooling improves memory performance and general cognition. We find evidence of heterogeneous effects by gender: the effects are stronger among women. We explore potential mechanisms, and we show that a more supportive social environment, improved health habits, and reduced stress levels likely play a critical role in mediating the beneficial effects of educational attainment on cognition among the elderly. (JEL J14, J24, I21, F63, N37)

[★]We are grateful to the research staff at the Harvard Center for Population and Development Studies who made the HAALSI data available to us and provided numerous insights based on their field experience implementing the survey. Plamen Nikolov gratefully acknowledges research support by the Economics Department at the State University of New York (Binghamton) and the Research Foundation for SUNY at Binghamton. We thank Charlotte Williams provided exceptional research assistance. We thank Matthew Bonci for constructive feedback and helpful comments. All remaining errors are our own.

[★]Contact information: Yeh: Department of Economics, Columbia University, 1022 International Affairs Building (IAB), 420 West 118th Street, New York, NY 10027, USA. E-mail: sy3012@columbia.edu; Nikolov: Department of Economics, State University of New York (Binghamton), LT 909, Binghamton, NY 13902, USA. E-mail: pnikolov@post.harvard.edu.

[a] State University of New York (at Binghamton)
[b] Columbia University
[c] IZA Institute of Labor Economics
[d] Harvard Institute for Quantitative Social Science
[e] Global Labor Organization

# I. Introduction

Economists have long been interested in the production function of human capital and have focused on cognitive skills because of their importance for economic decision-making. Historically, existing research primarily focuses on human capital accumulation in early childhood (Heckman, 2000; Heckman and Krueger, 2005). Considerably less evidence exists on the causes and consequences of human capital depreciation later in life. However, recent neuropsychological research shows that the human brain is malleable and susceptible to enhancement even in late adulthood. The economic and policy causes of cognitive performance in old age, an essential driver of well-being, productivity, and individual decision-making (Schimdt and Hunger, 2004), are not well understood. A better understanding of the causes of cognitive performance can shed light on the human capital production process and inform policies to enhance cognition in later life. Educational policies, in particular, can play a crucial role. We fill the research gap on this issue by examining the effect of schooling on cognition in the context of a developing country.[1,2,3]

We study the downstream effects of educational differences on old-age cognitive performance in South Africa. We use data from a panel survey, Health and Aging in Africa: A Longitudinal Survey in South Africa (hereafter, "HAALSI"). We use individual data from this survey, and we employ instrumental variable (IV) estimation techniques to estimate the average causal effects of educational differences on old-age cognition performance. We examine five cognitive outcomes: immediate episodic memory, delayed episodic memory, orientation, attention and speed, and executive functioning. These proxy measures capture various aspects of cognition, which play significant roles in distinct decision-making processes. Most notably, each proxy is vulnerable to age-related cognitive decline. For example, episodic memory is especially vulnerable to the aging process due to atrophy in the hippocampal region, the part of the brain controlling memory functions (Gorbach et al., 2017). Similarly, old-age neurological changes in the prefrontal lobe affect executive functioning (MacPherson et al., 2002). Reductions in attention capacity and deterioration in performance in the orientation domain are telltale signs of cognitive decline in aging adults (Smart et al., 2014; Guerrero-Berroa et al., 2009).

---

[1] Using data from high-income countries, Schneeweis et al. (2014) and Banks and Mazzonna (2012) show that more years of schooling can improve old-age cognitive performance
[2] Le Carret et al. (2003) and Mazzonna and Peracchi (2012) find positive associations between education and cognition in older adults.
[3] Schneeweis et al. (2014) analyze six European countries using data from the Survey of Health, Ageing and Retirement in Europe (SHARE). Banks and Mazzonna (2012) analyze cognitive performance in England using cognitive tests from the English Longitudinal Survey on Ageing.

Studying how educational differences affect old-age cognitive performance has critical economic implications. Because cognitive functioning determines how people process information, it is crucial for decision-making.[4] Increasingly, elderly individuals make complex financial, health, and long-term-care decisions, with significant economic consequences. Therefore, any cognitive decline among aging adults can adversely affect financial well-being (Korniotis and Kumar, 2011). Although financial literacy can buffer some of these adverse effects (DeLiema et al., 2020; Lusardi et al., 2014), the use of cognitive skills, such as coordination between the cognitive domains of executive functioning, attention and speed, and memory, is crucial for developing better financial literacy skills. Most low- and middle-income countries (LMICs) are undergoing an incisive demographic shift towards aging populations. Therefore, understanding how policies can affect cognitive performance among the elderly is paramount for developing effective interventions. South Africa makes for a compelling setting to study this issue as the country lacks intermediary market institutions to assist with financial decisions or health care.

Our analysis reveals several significant findings. First, based on the IV method, we show that an extra year of schooling has a protective effect on cognition by 0.2 standard deviations (SD). Based on the IV method, the effect size estimates imply that each additional year of schooling has a more substantial effect on the memory-related domains than on the other cognitive domains (i.e., executive functioning, orientation). This pattern persists when we use additional IVs. Second, we show that the positive effect on cognitive performance is more substantial among women. Among women, education improves cognitive performance by 0.06 SDs, an effect size larger than that for comparable men. Third, we shed light on the possible mechanisms underlying the relationship between educational attainment and old age cognitive performance. Specifically, we show that better-educated individuals tend to lead more active lifestyles, report residing in more hospitable local communities, maintain better health, engage in more social interactions using technology[5], and exhibit reduced stress levels. The specific mechanisms at play likely drive the discrepancy between the effect sizes differences associated with various cognitive domains.

---

[4] Mazzonna and Peracchi (2012) show that older people tend to underestimate their cognitive decline. Those who suffer from severe cognitive decline but are unaware of it are more likely to suffer sizeable financial wealth losses.

Our study contributes to our understanding of human capital in several ways. First, we provide clear evidence of downstream effects of how better educational attainment can confer substantial benefits on old-age cognition, a finding with significant implications in increasingly aging societies. We are also among the few studies to shed light on this issue using data from a developing country. Second, we show that the effect is more pronounced for domains responsible for higher-order cognition, such as memory-related domains. Finally, we explore the underlying mechanisms. We show that the cognition effects for better-educated people exhibiting better cognitive performance are likely due to three main channels: a more vigorous social life, living in more socially connected local communities, better health habits, and lower stress levels.

The remainder of this article is organized as follows. Section II provides background on the major cognitive domains, and we detail the data used in our analysis. Section III describes the empirical strategy. Section IV presents the results. Section V further discusses possible mechanisms and channels through which education may affect old-age cognitive abilities. Section VI describes various robustness checks. Section VII presents concluding remarks.

---

[5] HAALSI classifies technology use to encompass any of the following: phone (SMS messages), email, or the internet.

## II. Background

### A. Cognition

*Cognitive Domains.* Cognition generally comprises several domains. They are categorized based on the general process involved, such as language or memory, or the known locality in the brain responsible for particular cognitive functions. Although there is no consensus about a taxonomy of cognitive domains, previous research generally recognizes the following: memory, motor skills, attention, executive functioning, language, sensation, and perception (Harvey, 2019). Each of the domains works interdependently in enabling daily life functioning. For example, eating a meal involves the complex interaction between fundamental sensory, perceptual, and motor skill cognitive processes. Making decisions in old age, such as choosing between different risky assets, requires higher-order cognitive domains, such as attention, memory, executive functioning, and language.

We analyze five higher-order cognitive domains[6]: immediate episodic memory, delayed episodic memory, orientation, attention and speed, and executive functioning (Langa et al., 2020).[7] Immediate episodic memory describes the ability to recall immediately and mentally re-experience specific episodes from their past. Delayed episodic memory differs from immediate episodic memory concerning the timing of when the individual recalls the events. For instance, asking someone what they had just eaten a couple of minutes ago would test their immediate episodic memory, while inquiring about the contents of their dinner from two days ago would test their delayed episodic memory. The cognitive domain of orientation functions to keep one aware of three dimensions: time, place, and person. Attention and speed describe how an individual can selectively attend to specific information while ignoring irrelevant information. Finally, executive functioning encompasses higher-order cognition processes in generating plans, solving problems, and organizing structures that guide future actions. It also incorporates "basic sensory and perceptual processes" into the decision-making process (Harvey, 2019).

---

[6] Higher-order cognitive skills comprise the memory, attention, executive, language and perception domains.
[7] Langa et al. (2020) classified all five cognitive domains based on the approached adopted by the Health and Retirement Study (HRS) Harmonized Cognitive Assessment Protocol Project.

Recent neuroscience research underscores the importance of environmental factors, including educational attainment, for old-age cognition. For example, Greenwood and Parasuraman (2010) detect evidence that cognitive plasticity, the ability of the brain to make changes through neuronal growth and reorganization, is affected by formal education. People with better educational quality in childhood exhibit slower cognitive decline in old age (Colsher and Wallace, 1991). Zahodne et al. (2019) report evidence that education mediates the cognitive impact of white matter hyperintensities in the brain, negatively associated with memory performance.

**B.     Data and Sample Characteristics**

We analyze the HAALSI, a sister survey of the U.S.-based Health and Retirement Survey.[8] HAALSI is a longitudinal survey with follow-up surveys every three years. It primarily focuses on understanding the health trends of an aging population in rural South Africa.

The primary sample for the HAALSI is adults who are at least 40 years old and live permanently in the Agincourt region for 12 months before the 2013 Health and

---

[8] Data quality control was ensured through internal checks embedded within their data collection system, where analysis frequently checked data tables to monitor survey progress and data quality.

Demographic Surveillance System census.[9] The study site comprises 31 villages with a total population of 116,000 people. The baseline survey comprises 5,059 individuals and has a response rate of 86 percent (Gómez-Olivé et al., 2018). The gender split in the sample was approximately 46 percent male and 54 percent female.[10] Each survey interview lasted approximately three hours.

The survey collects demographic and socioeconomic characteristics, household composition, income and expenditures, and labor force participation. In addition to the extensive demographic information for each respondent, the survey collects information on childhood-related variables, paternal schooling, and the occupational status of parents. The HAALSI also captures individual health histories[11], physical examination outcomes[12], psychological well-being[13], and cognitive functioning. Focusing primarily on cognition outcomes, we use the baseline data collected between November 2014 and June 2015.

*Proxy Measures for Cognition in HAALSI.* We classify the cognitive domains tested in HAALSI using the categorization approach provided by Langa et al. (2020). Langa et al. (2020) describe the Harmonized Cognitive Assessment Protocol Project (HCAP), which seeks to measure better cognitive impairments of older adults in longitudinal studies from

---

[9] The young starting age was chosen for HAALSI because life expectancy at birth is low in South Africa (Gómez-Olivé et al., 2018). In addition, HAALSI aims to observe how diseases develop in individuals through middle age and how diseases affect individual health in later life (Gómez-Olivé et al., 2018).

[10] Ethnicity was not captured in HAALSI, as black Africans almost exclusively populated the survey's geographic coverage of Agincourt. Instead HAALSI collected data on nationality, which included the following: South African or Mozambique/Other.

[11] Individuals were asked about historical diagnoses of cardiovascular and metabolic conditions such as heart disease, hypertension, kidney disease, diabetes, tuberculosis, and HIV infection, in addition to questions about health care utilization and expenditure. They were also administered the International Physical Activity Questionnaire which asked them about the type and vigor of exercise they partook in during the week.

[12] The battery of physical examinations includes blood pressure, body mass index, height, weight, grip strength, waist-hip ratio, blood glucose, total cholesterol, and hemoglobin.

[13] HAALSI asks respondents about their psychological wellbeing using the Gallup (Well-being) Index, Center for Epidemiological Studies Depression scale, and a questionnaire about post-traumatic stress disorder.

high-, middle-, and low-income countries worldwide to allow for comparisons of outcomes across studies.

HAALSI elicits performance in the cognitive domains based on the HCAP classifications described in Langa et al. (2020). HAALSI tests the domain of immediate episodic memory by asking respondents to memorize a list of ten words. The enumerators then ask the respondents to list the words they remember immediately afterward. For the same domain, individuals are also asked to name the current president of South Africa. HAALSI tests executive functioning by asking respondents to fill in the blank in a numerical pattern.[14] Next, HAALSI tests the performance in the orientation domain by asking them the date— the day, month, and year. For the cognitive domain of attention and speed, the enumerators ask respondents to count to 20. If respondents count to 20 correctly, the score is marked as "correct." The delayed episodic memory test consists of the same word recollection examination as the immediate episodic memory. However, HAALSI surveyors ask the individuals to list the words they memorized after completing the number series and count to 20 tests.[15] For all responses to the cognitive tests, the interviewers are given individual audio-recorded answers. They score the accuracy of the response by noting if it is correct, incorrect, not known, or if the question is refused. All the cognitive tests occurred during the baseline HAALSI survey.

Based on the performance in each cognitive domain, we aggregate the information into a composite index. We perform this procedure for the immediate episodic memory domain: combining the first-round word recollection and remembered the president scores. We generate a composite index for memory items using individual test scores categorized under immediate episodic memory and delayed episodic memory.

---

[14] Respondents were read the following prompt: "I'm going to read you a series of numbers. There will be a blank number in the series that I read to you. I would like you to write down the numbers from left to right and then tell me what number goes in the blank based on the pattern of numbers. 2. . . 4. . . 6. . .BLANK." The correct solution is the number 8.

[15] Respondents were asked the following prompt: "A little while ago, I read you a list of words and you repeated the ones you could remember. Please tell me any of the words that you remember now."

Table 1 details the categorization of the cognitive domains and describes the composite cognitive measures.

[Table 1 about here]

*Other Study Outcomes.* In addition to the primary study outcomes, we use data on two additional variables – educational expenditures and the number of currently alive siblings –in the estimation strategy.

HAALSI collects data on household expenditures on education and training (e.g., tuition, school/training fees, books, uniforms, and other related expenses) in the past 12 months. The survey also collects data on the number of siblings who are alive for each survey respondent. We describe how we use these variables in the next section.

*Summary Statistics.* Table 2 reports the summary statistics for the study sample. The demographic characteristics mirror the profile of the South African rural community. However, high regional fertility rates have led to a slower demographic shift towards an aging population relative to the national population (Gómez-Olivé et al., 2018). Around 51 percent of the individuals are currently married. The average age is 61.74 years, and the average number of years of schooling is 4.12. There are, on average, five individuals in a typical household in the sample. The racial makeup is almost 100 percent black African (not shown in Table 2). Approximately 80 percent of the sample does not report formal employment. To obtain spending outcomes at the individual level, we divide the HAALSI-provided household expenditures values by household size. On average, individuals spend more money on education-related fees than healthcare and ceremonies in the past 12 months leading up to the interview date. The average individual expenses on education in the past 12 months is 395 ZAR (approximately USD 31). The spending variables are right-skewed, implying more people spend less than the average in the categories reported in Table 2.

The summary statistics also reveal significant gender discrepancies. Women, on average, attain fewer years of education than men. Women report lower employment rates than men. They are also less likely to be currently married (Gómez-Olivé et al., 2018).

[Table 2 about here]

## III. Empirical Strategy

### A. Ordinary Least Squares (OLS) Estimation

Our main objective is to estimate the effect of educational attainment on old-age cognitive performance. To quantify this effect, we start with a simple OLS specification of the following form:

$$y_{iv} = \beta_0 + \beta_1 E_{iv} + X\boldsymbol{\beta_2} + \gamma_v + \epsilon_{iv} \quad (1)$$

where $y_i$ is the standardized age-adjusted cognitive performance score for individual $i$ residing in village $v$. $E_{iv}$ measures years of education attained by individual $i$, $X$ is a vector that includes individual- and household-level control variables. $\gamma_v$ captures village fixed effects. We cluster at the village level.

We adjust the cognitive performance scores, which are our outcomes, for age-related effects. We adjust these scores using the residuals from linear regressions of the performance scores on age (Altonji et al., 2012). Since scores on cognitive performance tests generally correlate with the age and education level of the individual, the tests measure knowledge and not cognitive ability that is independent of specific knowledge (Altonji et al., 2012). Age-adjustment of the cognitive performance scores mitigates this issue.

The coefficient of interest, $\beta_1$, captures how an increase in a year of schooling affects the standardized age-adjusted cognitive performance. However, estimating this coefficient using an OLS is unlikely to yield unbiased estimates. Suppose we estimated (1) using an OLS: because the variation in educational attainment in (1) is not due to any exogenous variation, educational attainment will likely correlate with unobserved individual-level characteristics, such as community background characteristics, parental background, or parental care. Therefore, we attempt to address this endogeneity issue using an instrumental variable approach: we instrument educational attainment with several instrumental variables, detailed in the following sub-section.

### B.  Instrumental Variable (IV) Method

The primary estimation approach relies on an instrumental variable estimation. We instrument for individual educational attainment using several plausible instrumental variables. A valid instrument must meet two conditions: both affect years of schooling and cannot directly affect cognitive performance in old age.

Previous studies explore a range of potential instruments, including changes in schooling laws (Harmon and Walker, 1995), proximity to college (Card, 1993), and birth quarters (Angrist and Krueger, 1991). Supply-side factors are a possible source for identifying variation in demand-side parameters of schooling. Schooling expenditures can be a source of exogenous variation for years of schooling, assuming it is independent of taste and ability factors[16]. Given the specific setting of South Africa, Fiske and Ladd (2003, 2004) and Borkum (2012) find that enrollment at the secondary school level can be particularly sensitive to school fees.

Our primary instrument is the individual educational expenses in the past 12 months. The expenses, measured in South African ZAR[17], include education and training, tuition, school/training fees, books, uniforms, and hostel/meal charges related to education. We normalize the household education expenses by household size to proxy for individual education expenses. Using this variable, we can identify the so-called LATE: the average effect among the 'compliers', those whose educational attainment is affected by higher educational expenses. We estimate the LATE non-parametrically using a Wald estimator. The identifying assumption is that schooling expenditures affect cognitive performance only through one's educational attainment.[18]

---

[16] Grubb (1989), Rouse (1995, 1998), Kane (1994), and Kane and Rouse (1995) find that school expenditures and fees can be an important determinant of an individual's educational attainment.
[17] The 2015 average exchange rate was 1 ZAR = 0.0788 USD.

We note a robust first-stage estimation of the IV approach per the first-stage criterion described by Staiger and Stock (1997) and Stock and Yogo (2005). For educational expenses to serve as a plausible instrument for educational attainment to identify the effect of education on cognition, it must be the case that the educational expenses are validly excluded from the old-age cognitive performance production function. Such an exclusion is invalid if other variables correlate with cognitive performance and affect educational expenses. In particular, parental care or household-related inputs might fall into that category since such factors may affect cognition. In estimating the relationship between educational attainment and cognitive performance, we include controls for household characteristics and household parental care. Since we account for such factors, educational expenses would appear to be a plausible instrument.

## IV. Results: The Effect of Schooling Differences on Cognitive Decline

### A. Baseline Results

Table 3, Panel A reports the results based on the OLS. Columns (1) through (3) report the results on the individual memory measures. Column 4 reports the estimate for the composite measure, combining information from all memory items tested. In columns

---

[18] School fees might not be a valid instrument in a setting in which families move to certain communities because of their characteristics (if these characteristics are related to the school fees), or if school placement is nonrandom and related to fees (Pitt et al. 1993; Rosenzweig and Wolpin 1986).

(5) and (6), we report results for the cognition domains: executive functioning and attention/speed, respectively. Column 7 reports the coefficient for the composite index.

[Table 3 about here]

The results in Table 3 show that that schooling positively affects performance in all cognitive domains. The effect size is small: it ranges from 4 to 8 percent of an SD (the highest magnitude is on the delayed episodic memory measure). These estimates imply that an additional year of schooling is associated with a 0.08 SD increase in cognitive performance; similarly, an additional year of schooling increases memory performance by a 0.07 SD (for both immediate and delayed episodic memory).

**B.     Long-run IV Estimates of Educational Attainment on Cognitive Performance**

Next, we focus on the instrumental variable estimates, based on the education expense in the last 12 months variable. Panel B of Table 3 reports the 2SLS estimates.[19]

We point out several observations about the estimated coefficients. First, all coefficients are positive and statistically significant at the 1 percent level. The most negligible cognition gain associated with an additional year of schooling is the executive functioning domain: an increase of 0.100 SDs. In contrast, the delayed episodic memory exhibits the most sizable gain associated with an extra year of schooling: an increase of 0.230 SDs.

We report a substantial increase in the memory index and composite cognitive performance for the two composite measures. The 2SLS results imply that an additional school year improves general cognition and memory performance by 0.221 SDs and 0.223 SDs, respectively. The second noteworthy aspect of the 2SLS results is that they display higher returns to schooling than the OLS estimates from the same specifications. The difference in the OLS and IV estimates may be because the IV estimation relies on identifying variation from the so-called group of "compliers," which could have relatively high cognitive returns to education (Imbens and Angrist, 1994; Angrist, Imbens and Rubin, 1997). [20]

---

[19] We report standard errors clustered at the village level. The first stage Kleiberger-Paap Wald F-statistic is 81.831, implying a strong first stage.

It is important to note that the effect size estimates for cognitive domains related to memory are much larger than the other cognitive domains such as executive functioning and attention and speed. Two possible explanations can account for this difference. First, the HAALSI survey provides more granular detail in scoring memory performance (i.e., the word lists) than the latter two domains, which were only scored on a binary correctness basis. For instance, in the attention and speed cognitive domain, we cannot discern participants who count to 19 versus those who only count to 5. More detailed survey results can provide more insight into the true variation of executive functioning and attention and speed cognitive domain performance. However, such was not the case in our data. Thereby, it could be that HAALSI captures the memory performance more accurately than the other domains. Second, the difference in the effects of schooling on the various cognitive domains could be because of different mechanisms mediating the effects. For example, in the results presented in Section V, we find that an extra year of education reduces overall stress. Researchers link stress to memory decline in aging individuals[21], which could likely account for the higher effect size associated with the memory outcomes (immediate episodic and delayed episodic) than those of the other cognitive domains (McEwen, 2000; Lupien et al., 2009; Sandi, 2013). We also explore (in Section V) explore for possible mechanisms through which schooling affects cognition.

---

[20] If schooling and ability are positively corelated, then the OLS estimates of how schooling affects cognition will be *greater than* the 2SLS estimates. This pattern will hold up when both schooling and cognition are measured contemporaneously. In our analyses cognition is measured decades after schooling and cognitive performance can deteriorate over time and this decline could occur in such a manner that individuals with high cognitive endowments exhibit faster decline rates than those with lower cognitive endowments. Given that HAALSI does not collect data on individual cognitive abilities at different stages of life, we are unable to address this particular issue.

[21] The underlying neurological reasons for why this phenomenon happens is discussed in Section D.

Next, we compare our findings to previous research on the same issue based on data from high-income countries. Our estimates for the effect of an additional year of schooling on memory are considerably greater than those of Schneeweis et al. (2014), which reports that one additional year of schooling increases performance in immediate memory and delayed memory by roughly 0.14 and 0.17 SDs, respectively.[22] However, when comparing the effect size estimates of schooling on memory, our estimates are lower than those in Banks and Mazzonna (2012), which reported effect sizes of around 0.5 SDs for males and 0.4 SDs for females.[23] Our findings on the effect sizes of schooling on memory[24] thus are within the range of effect sizes established using the results of the previous two studies. We conclude that the size of the effect of schooling on memory does not differ significantly from those quantified in a high-income setting. In the context of other studies, our results imply that education exerts similar effects on memory performance in old age across societies of varying wealth and economic development progress. Education is a universally vital input in the production function of memory performance in old age.

We turn to the effect sizes of the other cognitive domains. Schneeweis et al. (2014) focus on numeracy cognitive skills using a numeracy test most closely resembling the number series performance metric, testing executive functioning in HAALSI. Schneeweis et al. (2014) find a negative but statistically insignificant effect size of -0.013 SDs for executive functioning and a statistically insignificant effect size of -0.007 SDs for orientation. In contrast, we find a more sizeable positive effect size of 0.141 SDs for orientation and 0.100 SDs for executive functioning. On the other hand, our effect size on executive functioning is smaller than Banks and Mazzonna (2012); the study finds large and significant effects of schooling reform on executive functioning upwards of 0.6 SDs. Thus, our estimates reveal that even in the developing economy setting, the effect of education on executive functioning is similar to the one previously observed with data from high-income settings (Banks and Mazzonna 2012). There are no other studies using data on the memory proxy, so we cannot benchmark the effect size to previous research.

---

[22] The cognitive test for immediate and delayed memory analyzed by Schneeweis et al. (2014) also consisted of asking respondents to memorize and recall a list of ten words.
[23] Banks and Mazzonna (2012) did not report effect size estimates for the sample with men and women pooled together, and thus we fall short of a direct comparison of the effect sizes for memory and executive functioning.
[24] Note that we are referencing the coefficient estimate on the composite memory index, which is Table 3 Column (4).

### C.    Heterogeneous Treatment Analysis

Next, we examine how the effect of an additional year of schooling on cognitive performance differs by gender.

Recent studies document gender differences regarding the onset and the severity of the cognitive decline.[25] Gender differences may be more pronounced in the South African setting, particularly in rural regions.[26] In most rural and traditional regions of South Africa, men usually act as the head of the household and make most of the household decisions, which can reduce the exposure of women to the various transmission mechanisms[27] through which more education can positively affect cognitive performance (Cheteni et al., 2019). The patriarchal society also imposes cultural and social norms, negatively affecting women's economic situation (Cheteni et al., 2019).[28]

---

[25] Using data from the United States, Levine et al. (2021) find that women experienced a quicker decline in global cognitive abilities and executive function than men, but not memory. Wang et al. (2020) examine cognitive performance data drawn from aging individuals in rural regions of China and discover a higher prevalence of cognitive impairment in women than men. In contrast, a study conducted by Barnes et al. (2003) utilizing longitudinal data from elderly Catholic participants in the Religious Orders Study finds that the patterns of cognitive decline and the development of Alzheimer's Disease[25] are similar among men and women in old age. Nevertheless, the significance of our results may indeed provide insights for potential policies aimed at improving the cognitive well-being of men and women in old age.

[26] Banks and Mazzonna (2012) and Schneeweis et al. (2014) explore gender differences with data from high-income countries. The former finds long-term positive effects for men only, while the latter find larger effects among men than women.

[27] We explore mechanisms related to social interactions, social environments, labor outcomes, stress, health habits, and entertainment— all of which may be impacted if in fact the men of the household, rather than the women, are making the household decisions.

[28] Low socioeconomic status is a predictor of cognitive decline in aging individuals (Koster et al. 2005).

Therefore, it is critical to evaluate whether experiencing differential schooling affects cognition in old age in a setting laden with gender discrimination and biased societal norms.

We estimate the heterogeneous effects using the following specification:

$$y_{iv} = \beta_0 + \beta_1 E_{iv} + \beta_2 E_{iv} \times Female_{iv} + \mathbf{X}\boldsymbol{\beta_3} + \gamma_v + \epsilon_{iv} \quad (2)$$

We define all variables as in the primary specification (1). The coefficient $\beta_2$. captures the heterogeneous effects by gender.

Table 4 reports estimates of the interaction term, $\beta_2$, based on equation (2). Panel A reports estimates based on the OLS; Panel B reports estimates based on the 2SLS. The estimates of the interaction term $\beta_2$ are reported in the third sub-row.

[Table 4 about here]

Based on the OLS estimation, the interaction term is positive for all cognitive domains and statistically significant for most cognitive domains. The significant coefficient implies that higher educational attainment improves women's cognitive performance more than men.

We turn to the heterogeneous treatment analysis based on the 2SLS estimation approach.[29] The two-stage least squares estimates (reported in Panel B) reveal considerably less statistically significant gender differences in old-age cognitive performance. We detect significant and positive coefficients on three individual outcomes (immediate episodic memory in Column 1, orientation in Column 3, and attention and speed in Column 6). Compared to men, an extra year of schooling improves women's performance by 0.058 SDs, 0.057 SDs, and 0.097 SD on immediate episodic memory, orientation, and attention and speed, respectively. Only the point estimate for attention and speed is significant at the 1-percent level. The coefficient estimate for the composite cognition index is positive at 0.062 SDs and significant (at the 10 percent level) and not significant at any level for the composite memory index. In sum, there are minimal statistically significant gender differences in how schooling affects cognitive performance.

---

[29] The first stage Kleiberger-Paap Wald F-statistic is 27.831, implying a strong first stages.

Our results contrast with the effects noted in Banks and Mazzonna (2012): an additional year of education does not affect the cognitive performance of men and women differently. Our effect size estimates differ from Schneeweis et al. (2014), which find more substantial protective effects of schooling on cognition for males. However, we reach a similar conclusion to Schneeweis et al. (2014) in that the coefficient estimates are not statistically significant between men and women. The difference in the quality of education and gender bias in the schooling system in Europe[30] and South Africa may account for the apparent differences we observe in our results and those of the other two studies. HAALSI does not provide further details into childhood education quality or data that could account for perceived gender inequality in the education system. Thus, we are limited in further exploring the underlying reason for the differences.

## V. Discussion

We next discuss how our main results can be rationalized in light of different channels. We explore several possible mechanisms: health behaviors, stress, social interactions, the social environment, occupational choice, and participation in cultural activities.

### A. Possible Mechanisms

---

[30] Again, Banks and Mazzonna (2012) studies education effects using data from England while Schneeweis et al. (2014) utilizes data from six different European countries.

*Social Interactions.* The first possible channel via which education could positively affect cognitive performance is through a higher rate of social interactions. Although the channel is not well-understood in the literature, the reasons behind this protective effect may be primarily biological. The neurobiological theory underpinning this particular mechanism highlights the importance of enrichment-derived hippocampal protection. In other words, more social engagements could drive higher exposure to a variety of stimuli, which in turn drives higher neural activity. The higher neural activity can result in the formation of a cognitive reserve. An adept cognitive reserve provides the brain with alternative cognitive strategies to compensate for age-related impairments and preserve cognitive functioning (Fratiglioni et al., 2000). Furthermore, although aging leads to a drop in the birth and survival of new neurons in this brain region, other factors could slow down this process. For example, exposure to a richer set of environmental factors and higher engagement in social activities can partially alleviate this loss, primarily by enhancing the survival of newly born neurons (Kempermann et al., 2002). Several empirical studies document this relationship (Zunzunegui et al., 2003; Fratiglioni et al., 2000).

HAALSI collects data on how frequently individuals interact with their children and people they consider close to them. The survey collects expenditure data on communication-related purchases in the past 30 days. It also collects data on the type and frequency of social interactions, such as visiting children and interacting with close contacts using technology.[31] Other variables collected in the survey include participation in cultural and entertainment events and spending habits on entertainment and ceremonial items.[32]

---

[31] The survey classifies technology as using the phone, by SMS, through email or the internet.
[32] HAALSI specifically classifies entertainment spending as spending on games, table pool stadium soccer matches and attending bashes in the past 30 day. The questionnaire lists any spending includes spending on festivals and religious ceremonies such as festival clothes gifts, foods, etc. in the past 12 months as ceremony spending. All spending variables are in South African ZAR.

*Community Environment.* A second channel relates to the features of the immediate social environment that allow for better assimilation and integration within the community. In particular, higher-quality neighborhoods with better physical infrastructure and better social capital could confer additional benefits on cognitive performance, particularly in old age. Neighborhoods of higher socioeconomic status may promote cognitive function and buffer cognitive decline. Higher density of physical resources (recreational centers, parks, and walking paths) and social and institutional resources (i.e., libraries or libraries community centers) typically characterize such communities. These amenities can promote protective health behaviors (e.g., physical activity) and facilitate mental stimulation (e.g., social interaction and cognitive activities such as reading and playing games). People who exhibit higher levels of community social integration also have an associated lower probability of cognitive decline (Zunzunegui et al., 2003). A lack of social support is another factor associated with cognitive decline in the elderly (Dikinson, 2011).

HAALSI collects data on neighbors' willingness to help, village security, the general trustworthiness of the people in their village, and various proxies of social reciprocity. We explore these outcomes in our mechanism analysis.

*Stress.* The third channel could work through reducing stress levels. More educated people can sort into occupations and environments characterized by reduced levels of stress. Repeated stress causes the shortening and debranching of neurons in the hippocampus and reduces hippocampal volumes, the brain region primarily responsible for memory functions (McEwen, 2000; Lupien et al., 2009). Sandi (2013) finds that although mild stress improves cognitive function, high-intensity stress can chronically damage memory formation. Existing studies show that higher stress levels lead to faster cognitive decline among the elderly (Turner et al., 2017; Nilsen et al., 2014).

We use data from the HAALSI to assess changes in stress levels based on the educational level. The survey collects information on whether respondents felt stressed, worried, sad, angry, or happy.[33] To avoid a possible multiple-inference problem, we generate a summary index based on all available outcomes in the specific mechanism domains we explore (based on the approach noted in Anderson, 2008).

*Health Behaviors.* Better health behaviors can also reduce the risks of cognitive decline in old age. Several studies point to beneficial cognitive impacts through reducing tobacco smoking (Anstey et al., 2007), lower alcohol consumption (Topiwala et al., 2017), and improved diets (Morris et al., 2006).[34] The HAALSI collects data on current smoking, drinking habits, fruits/vegetable consumption, and healthcare spending in the last 12 months by the individual.[35]

*Labor Market Outcomes.* Finally, early retirement can impede cognitive health and accelerate cognitive decline (Rohwedder and Willis, 2010; Celidoni et al., 2017, Bonsang et al., 2012; Nikolov and Adelman, 2020). To address this potential mechanism, we use HAALSI data on whether a survey participant currently reports being employed or not.

## B. Supporting Evidence

The above channels are not mutually exclusive. Given the HAALSI data limitations, we cannot take on a particular interpretation against the others. In what follows, we investigate the plausibility of some of the hypothesized mechanisms using the existing survey data. We estimate the following specification:

$$y_{iv} = \beta_0 + \beta_1 E_{iv} + \boldsymbol{X}\boldsymbol{\beta_2} + \gamma_v + \epsilon_{iv} \quad (3)$$

---

[33] The variables are therefore coded in a binary fashion— 1 if the respondents did have the particular emotion and 0 otherwise.
[34] This is also classified as the so-called Mediterranean diet.
[35] Healthcare expenses consists of fees spent on doctors, nurses, dentists, hospitals, clinics, medicines, bandages, medical supplies, chemist and pharmacy purchases, traditional healers, and other medical costs.

where $y_i$ is the proposed mechanism for individual $i$ residing in village $v$, and all other variables are the same as in specification (1). We cluster errors at the village level. $\beta_1$ captures the effect of an extra year of schooling on the proposed transmission mechanism variable. To avoid possible multiple hypothesis testing issues due to the multiplicity of the testing framework, we also report the Westfall-Young (WY) adjusted p-values based on a correction procedure developed by Westfall et al. (1993).[36]

Table 5 reports the estimations based on (3) for the first two mechanism outcomes: social interactions and the social environment. The social interaction variables are binary indicators: 1 if the social interaction occurs at least once per week and 0 otherwise. Similarly, the social environment variables are binary indicators: 1 if the respondent strongly agrees with the social perception noted and 0 otherwise. Column (5) reports the estimates for the expenditure variable on communication-related items (in ZARs) in the past 30 days. Columns 6 and 11 report estimates for the composite indices for social interactions and social perceptions. Panel A reports estimates based on a maximum likelihood procedure. Panel B reports estimates from a linear probability model, and Panel C reports the IV estimates.

[Table 5 about here]

The estimates reveal that schooling positively affects the likelihood that individuals interact with their children or close friends using technology at least once per week. Likewise, we show a positive relationship between education and the individual perception of better social cohesion and more hospitable local community surroundings. The IV estimates bolster the pattern for these two outcomes. More schooling also increases individual communication expenditures by 23.1 percent, significant at the 1-percent level per its WY p-value. Based on the WY adjusted p-values, we detect that most mechanism proxies for social interactions are statistically insignificant. In contrast, those for the social environment are significant at the 10-percent level.

---

[36] Westfall and Young proposed step-down procedures that use permutation calculations to estimate dependence relationships. The procedure produces an adjusted correction statistic for multiple testing under possible dependence.

Turning to the effect of education on the composite mechanism indices, we observe that the 2SLS estimates for both outcomes are positive. However, only the perception of the social environment index is positive and statistically significant. An increase in a year of schooling leads to a -0.016 SDs increase in the social interaction index (not significant). The social environment index increases by 0.103 SDs (significant at the 1-percent level) by one more year of schooling. The latter result confirms that people with more education report better social aspects of their living environments.

Table 6 presents the mechanism analysis for employment, health behaviors, entertainment, and stress. Columns 1-3 report the estimates based on specification (3)—the variables in these columns (employment status, smoking, consuming alcohol) are binary-defined. We proxy stress levels using binary indicators: 1 if the survey participant experienced the particular emotion the day before the interview and 0 if not. Healthcare spending (reported in Column 4) includes spending on doctors, nurses, hospitals, or medicines in the past 12 months. We report results on fruits/vegetable spending in the last 30 days..[37] HAALSI asks respondents to report their ceremony spending on entertainment and cultural participation, including festivals and religious ceremonies such as festival clothes, gifts, or food items in the past 12 months. Entertainment spending includes games, table pool, stadium soccer matches, and attending bashes in the past 30 days. Note that all expenditure variables are denominated in ZAR.[38] In Table 6, Panel A presents the maximum likelihood estimates, Panel B reports OLS estimates, and Panel C presents the IV estimates.

[Table 6 about here]

---

[37] The full list of fruits and vegetables: apple, banana, guava, avocado, orange, naartjie, lemon, grape, raisin, melon, grapefruit, pineapple, mulberry, wild fruit, butternut, carrot, spinach, muroho, cabbage, broccoli etc.
[38] Again, note that we divide the household expenditure values by the household size to obtain a proxy for individual spending levels.

The IV estimates reveal that an extra year of schooling increases the probability of working and leads to decreased smoking and alcohol consumption. In addition, an extra year of schooling increases health spending by 73.6 percent over 12 months (significant at the 1-percent level based on the WY p-value) and a 1.7 percent increase in spending on fruits and vegetables in 30 days (not significant at any level based on its WY p-value). Overall, we detect a positive impact on the composite health behaviors index by 0.054 SDs, indicating the education improves health behaviors in old age albeit a small effect size.

Columns (7) and (8) report the findings on the entertainment and cultural participation proxies. We see that a year of schooling leads to a 76.5-percent increase in spending on ceremonies in 12 months (significant at the 1-percent level based on the WY p-value) and a 9.3-percent increase in entertainment expenditures in the last 30 days (significant at the 10-percent level based on the WY p-value). Turning to the proxy variables for stress levels, we detect that the IV estimates reveal that an extra year of schooling improves performance on the composite stress index by 0.144 SDs. We detect evidence that education lowers most individual stress proxies; the proxy for anger is significant at the 1-percent level.

In summary, and based on the composite index variables, we detect that an extra year of schooling (1) lowers stress, (2) increases positive social perceptions, (3) increases social participation in cultural activities, and (4) improves healthy behaviors. Based on the effect size estimates, there is suggestive evidence that two main channels most likely mediate the protective effects of education on cognitive performance in old age: reduced stress levels and creating positive social perceptions.

## VI.  Robustness Checks

This section presents additional robustness analyses to corroborate the magnitude of the estimates based on the IV estimation procedure. We employ two additional instruments for these exercises: the number of living siblings[39] and a binary indicator for a positive amount of money spent on education for educational attainment based on the IV estimation procedure outlined by equation (2). Table 7 reports the results. Panel A reports the results using the number of siblings instrument; Panel B reports the results using the money spent on education binary indicator.

[Table 8 about here]

The estimates reported in Table 8 are similar to those based on the primary instrument reported in Table 3. In terms of the composite cognition index, the coefficients based on the two alternative instruments yield slightly larger coefficient estimates: 0.299 SDs for the number of siblings and 0.323 SDs for the money spent on education indicator. However, all coefficients reported in column (7) are similar to those reported in Table 3, and they are statistically significant at the 1-percent level. Similarly, the effect size estimates associated with the cognitive outcomes –immediate episodic memory, orientation, attention and speed, and the composite memory index— are slightly larger based on the two alternative instruments than those based on the primary IV. The baseline estimates (reported in Table 3) for executive functioning (0.100 SDs) fall between the estimates based on the number of siblings IV(0.140 SDs) and the money spent on education indicator IV (0.066 SDs). The same pattern holds for the delayed episodic memory domain. The estimated effect (i.e., 0.230 SDs) from Table 3 falls in the range of the estimates based on the additional IVs. The effect size based on the number of siblings instrument is 0.213 SDs; the point estimate based on the money spent on education indicator is 0.330 SDs. In sum, based on the two alternative instruments, we find evidence bolstering the pattern and magnitude of the primary estimation results reported earlier.

---

[39] The results using the number of siblings currently alive may violate the strict exogeneity condition in that siblings may affect an individual's cognitive development at a young age that is independent of educational attainment (Azmitia and Hesser, 1993).

Next, we conduct the 2SLS analysis split among subsamples of individuals born in South Africa and those who were not. We do so out of concern that the latter tend to have, on average, lower attained years of schooling, and their educational experience may be vastly different from those born in South Africa.[40] The results presented in Table A.1 for the South Africa-born subsample are slightly lower than those of the main results. We find positive and larger estimate effects for the foreign-born subsample.[41]

We also conduct an additional exercise in which we limit the influence of extreme values— particularly those that attained much higher years of education than others – on the estimation procedure. We re-estimate the IV model omitting the individuals who had completed university. We report the results in Table A.2. The effect sizes based on the winsorized analysis are slightly greater than those in the main results, by about 0.02 SDs for each cognitive domain.[42] Therefore, excluding outliers produces results consistent with the ones based on the whole sample.

## VII. Conclusion

While extensive literature has focused on the effects of education on cognitive performance in early childhood, there are considerably fewer studies examining how educational differences influence cognitive performance in later life. This study investigates the consequences for better educational attainment on cognitive performance using a data sample from South Africa. Notably, it is the first study to consider how education affects cognition in old age using data from a developing country. Our study provides insights into the long-term consequences of higher human capital into adulthood.

---

[40] 70 percent of the sample are born in South Africa. Those born in South Africa on average completed 5.32 years of schooling compared to 1.43 years of schooling for those born outside of South Africa.
[41] However, the 1st Stage Kleiberger-Paap Wald F-statistic of 8.519 for the analysis of the foreign sub-sample no longer passes the criterion of a strong instrument.
[42] We are primarily concerned that these individuals who achieved the most years of education are responsible for the strong first-stage. Even after omitting them from the analysis, the 1st stage Kleiberger-Paap Wald F-statistic is 62.72, which still indicates a very strong instrument.

We find that an additional year of schooling increases cognitive performance in several domains, including immediate episodic memory, delayed episodic memory, orientation, executive functioning, and attention and speed. An extra year at school increases aggregate cognitive performance across all five domains by about a quarter of an SD. We investigate possible channels to interpret our main results. There is suggestive evidence that two main channels most likely mediating the protective effects of education on cognitive performance in old age are: reduced stress levels and creating positive social perceptions.

The policy implications from our analysis are clear— more schooling can lead to significant downstream benefits for the elderly. Education can improve better cognitive performance for five critical cognition domains. Most prior analyses focus predominantly on the contemporaneous impacts of educational attainment or the impacts on children in early childhood. The focus on early life can understate the lifetime returns to education. The consequences of reduced cognitive performance in old age can affect well-being, work performance and could significantly improve individual decision-making. Since the existing literature is still relatively small, there is a considerable need for additional research to shed more light on how better cognitive performance in later life can improve the welfare of aging adults. Furthermore, future research should better understand the mechanisms underlying the relationship between educational attainment and cognition in late adulthood.

**Table 1:** Cognitive Domains in HAALSI

| Cognitive Domain | Domain Description | HAALSI Test |
|---|---|---|
| Immediate Episodic Memory (MemIE) | Ability to immediately recall and mentally re-experience specific episodes from the past | CERAD Word List Learning and Recall – Immediate Score, Remembered the President |
| Delayed Episodic Memory (MemDE) | Ability to recall and mentally re-experience specific episodes from the past after a timed delay | Delayed Word Recollection |
| Orientation | Ability and awareness of three dimensions: time, place, and person | Date Recollection |
| Executive Functioning | Allows one to generate plans, solutions to problems, or organizing structures that guide future actions | Number Series |
| Attention and Speed | Ability to selectively attend to specific information while ignoring irrelevant information | Count to 20 |
| Aggregate Memory | A composite index of the test scores in the MemIE and MemDE domains | |
| Aggregate Cognition | A composite index of test scores across all domains | |

**Table 2:** Summary Statistics

|  | Mean (St Dev) |
|---|---|
| *Socioeconomic Characteristics* | |
| Gender (=1 if Female) | 0.536 (0.499) |
| Age | 61.738 (13.062) |
| Education (in years) | 4.147 (4.701) |
| Currently Married (= 1 if Yes) | 0.509 (0.500) |
| Father Attended School (= 1 if Yes) | 0.144 (0.351) |
| Household Size | 5.330 (3.342) |
| # Siblings Alive | 3.234 (1.970) |
| Born in South Africa (= 1 if Yes) | 0.700 (0.459) |
| Currently Employed (= 1 if Yes) | 0.201 (0.401) |
| Consumes Alcohol (= 1 if Yes) | 0.231 (0.422) |
| Currently Smokes (= 1 if Yes) | 0.091 (0.288) |
| *Individual Spending Outcomes* | |
| Money Spent on Education (in South African ZAR)[a] | 395.283 (2222.408) |
| Any money Spent on Education (= 1 if Yes)[a, c] | 0.427 (0.495) |
| Money Spent on Healthcare (in South African ZAR)[a] | 168.621 (723.109) |
| Money Spent on Communications (in South African ZAR)[b] | 30.273 (45.359) |
| Money Spent on Ceremonies (in South African ZAR)[a] | 93.340 (437.064) |
| Money Spent on Entertainment (in South African ZAR)[b] | 3.359 (67.647) |
| Money Spent on Fruits and Vegetables (in South African ZAR)[b] | 21.274 (24.450) |
| *Cognitive Outcomes (Non-Standardized)* | |
| Words Recalled (1st Round) | 4.461 (1.768) |
| Words Recalled (2nd Round) | 4.137 (1.868) |
| Number Series | 0.643 (0.479) |
| Remembered the Date | 2.303 (1.102) |
| Remembered the President | 0.829 (0.376) |
| Count to 20 Correctly | 0.782 (0.413) |
| Observations | 5,059 |

*Notes:* Data source is Health and Aging in Africa: A Longitudinal Study— Baseline Survey (2014-2015). All spending items are in ZAR; The avg 2015 exchange rate was 1 ZAR = 0.0788 USD. To obtain individual spending outcomes, we divide the HAALSI-provided household expenditures by household size. (a) Spending behavior time frame— the past 12 months. (b) Spending behavior time frame— the past 30 days. Money spent on education consists of tuition, school/training fees, books, uniforms, and other related expenses (including hostel/meals and charges related to education). (c) We report the indicator of money spent on education as it serves as an instrument in our 2SLS analysis. For Words Recalled (1st Round and 2nd Round), respondents were asked to recall the words they were asked to memorize. The maximum score is 10 words and minimum score is 0 words. Number series is a binary variable that is 1 if correct and 0 if incorrect— individuals were asked to complete the following numerical pattern: 2, 4, 6, __. The correct solution is 8. Remembered the Date asks participants to state the current year, month, and day. The maximum value is 3 if the respondent answers all three aspects of the date correctly, 2 if only two items are correct, and etc. The minimum value is 0 if all are incorrect. Survey respondents are asked to state the current president of South Africa for the Remembered the President variable, which is 1 if correct and 0 if incorrect. For Count to 20 Correctly, participants are asked to count to 20. The variable is 1 if correct and 0 if incorrect.

**Table 3:** Effects of Schooling on Performance in Various Cognitive Domains (HAALSI)

|  | Immediate Episodic Memory | Delayed Episodic Memory | Orientation | Aggregate Memory Index | Executive Functioning | Attention and Speed | Aggregate Cognition Index |
|---|---|---|---|---|---|---|---|
|  | (1) | (2) | (3) | (4) | (5) | (6) | (7) |
| *Panel A: OLS* | | | | | | | |
| Years of Schooling | 0.058*** | 0.063*** | 0.061*** | 0.067*** | 0.064*** | 0.044*** | 0.079*** |
|  | (0.005) | (0.005) | (0.006) | (0.005) | (0.005) | (0.004) | (0.005) |
| $R^2$ | 0.164 | 0.116 | 0.179 | 0.166 | 0.134 | 0.192 | 0.249 |
| Observations | 5,059 | 5,059 | 5,059 | 5,059 | 5,059 | 5,059 | 5,059 |
| *Panel B: 2SLS* | | | | | | | |
| Years of Schooling | 0.175*** | 0.230*** | 0.141*** | 0.223*** | 0.100*** | 0.145*** | 0.221*** |
|  | (0.025) | (0.050) | (0.028) | (0.040) | (0.028) | (0.033) | (0.028) |
| 1st Stage F-stat | 81.831 | 81.831 | 81.831 | 81.831 | 81.831 | 81.831 | 81.831 |
| $R^2$ | -- | -- | 0.093 | -- | 0.117 | 0.056 | -- |
| Observations | 5,059 | 5,059 | 5,059 | 5,059 | 5,059 | 5,059 | 5,059 |

*Notes:* Data source is Health and Aging in Africa: A Longitudinal Study. Each coefficient and standard error in each cell represents a different regression. Panel A reports OLS estimates of years of education on cognitive performance; Panel B reports 2SLS estimates using log household school expenditures as the instrument. The reported 1st Stage F-stat is the Kleiberger-Paap Wald F-statistic. Aggregate Memory is the first principal component of all memory-related test scores. Aggregate Cognition is the first principal component of all cognition test scores. Each cognitive test measure is age-adjusted. Each regression contains in its specification: indicators for gender, age, indicators for marital status, indicators for whether the individual's father attended any school, indicators for country of origin, dummies for the villages the respondents reside in (they are encoded anonymously for a total of 27 villages). Standard errors are clustered at the village level.

*** Significant at the 1 percent level. ** Significant at the 5 percent level. * Significant at the 10 percent level.

**Table 4:** Heterogeneous Effects of Schooling on Performance in Various Cognitive Domains (HAALSI)

|  | Immediate Episodic Memory | Delayed Episodic Memory | Orientation | Aggregate Memory Index | Executive Functioning | Attention and Speed | Aggregate Cognition Index |
|---|---|---|---|---|---|---|---|
|  | (1) | (2) | (3) | (4) | (5) | (6) | (7) |
| *Panel A: OLS* | | | | | | | |
| Years of Schooling | 0.049*** | 0.061*** | 0.035*** | 0.063*** | 0.049*** | 0.012*** | 0.056*** |
|  | (0.006) | (0.005) | (0.006) | (0.006) | (0.006) | (0.004) | (0.006) |
| Female | -0.157*** | 0.020 | -0.500*** | -0.037 | -0.354*** | -0.632*** | -0.434*** |
|  | (0.039) | (0.042) | (0.050) | (0.038) | (0.040) | (0.060) | (0.048) |
| Female × Years of Schooling | 0.017*** | 0.005 | 0.054*** | 0.009 | 0.032*** | 0.068*** | 0.047*** |
|  | (0.006) | (0.006) | (0.005) | (0.006) | (0.004) | (0.007) | (0.005) |
| $R^2$ | 0.166 | 0.116 | 0.194 | 0.166 | 0.140 | 0.217 | 0.261 |
| Observations | 5,059 | 5,059 | 5,059 | 5,059 | 5,059 | 5,059 | 5,059 |
| *Panel B: 2SLS* | | | | | | | |
| Years of Schooling | 0.148*** | 0.238*** | 0.114*** | 0.215*** | 0.095*** | 0.101*** | 0.193*** |
|  | (0.029) | (0.056) | (0.027) | (0.046) | (0.027) | (0.027) | (0.026) |
| Female | -0.251* | 0.229 | -0.459*** | 0.030 | -0.241* | -0.686*** | -0.399*** |
|  | (0.133) | (0.172) | (0.101) | (0.157) | (0.127) | (0.138) | (0.133) |
| Female × Years of Schooling | 0.058** | -0.017 | 0.057** | 0.018 | 0.011 | 0.097*** | 0.062* |
|  | (0.029) | (0.040) | (0.023) | (0.035) | (0.029) | (0.033) | (0.032) |
| 1st Stage F-stat | 27.831 | 27.831 | 27.831 | 27.831 | 27.831 | 27.831 | 27.831 |
| $R^2$ | -- | -- | 0.105 | -- | 0.120 | 0.069 | -- |
| Observations | 5,059 | 5,059 | 5,059 | 5,059 | 5,059 | 5,059 | 5,059 |

*Notes:* Data source is Health and Aging in Africa: A Longitudinal Study. Each coefficient and standard error in each cell represents a different regression. Panel A reports OLS estimates of years of education on cognition and cognitive decline; Panel B reports 2SLS estimates using log household school expenditures and log household school expenditures interacted with gender as the instruments. The reported 1st Stage F-stat is the Kleiberger-Paap Wald F-statistic. Aggregate Memory is the first principal component of all memory-related test scores. Aggregate Cognition is the first principal component of all cognition test scores. Each cognitive test measure is age-adjusted. Each regression contains in its specification: indicators for gender, age, indicators for marital status, indicators for whether the individual's father attended any type of school, indicators for country of origin, dummies for the villages the respondents reside in (they are encoded anonymously for a total of 27 villages). Standard errors are clustered at the village level.

*** Significant at the 1 percent level. ** Significant at the 5 percent level. * Significant at the 10 percent level.

**Table 5:** Mechanisms of the Effects of Schooling on Possible Mediating Variables

| | Social Interactions | | | | | | Social Environment | | | | | Labor |
|---|---|---|---|---|---|---|---|---|---|---|---|---|
| | Visit Children[a] | Children w/ Tech[a] | Contact w/ Close Contacts[a] | Close Contacts w/ Tech[a] | Log Spending Comms.[b] | Social Interactions Index | Helpful Neighbors[c] | Village Trust[c] | Helpful Village[c] | Village Security[c] | Social Environment Index | Works (=1 if Yes)[d] |
| | (1) | (2) | (3) | (4) | (5) | (6) | (7) | (8) | (9) | (10) | (11) | (12) |
| *Panel A: Probit* | | | | | | | | | | | | |
| Years of Schooling | -0.006 | 0.021*** | 0.006 | 0.045*** | -- | -- | 0.012** | 0.006 | 0.009* | 0.008 | -- | 0.055*** |
| | (0.005) | (0.006) | (0.007) | (0.007) | | | (0.005) | (0.005) | (0.005) | (0.005) | | (0.006) |
| W-Y p-value | [0.70] | [0.05]* | [0.70] | [0.00]*** | -- | -- | [0.24] | [0.37] | [0.37] | [0.37] | -- | -- |
| Observations | 5,059 | 5,059 | 5,059 | 5,059 | -- | -- | 5,059 | 5,059 | 5,059 | 5,059 | -- | 5,059 |
| *Panel B: OLS* | | | | | | | | | | | | |
| Years of Schooling | -0.002 | 0.006*** | 0.001 | 0.007*** | 0.062*** | 0.014*** | 0.004** | 0.002 | 0.003* | 0.003* | 0.007** | 0.018*** |
| | (0.002) | (0.002) | (0.001) | (0.001) | (0.004) | (0.005) | (0.002) | (0.001) | (0.002) | (0.001) | (0.003) | (0.001) |
| W-Y p-value | [0.70] | [0.06]* | [0.73] | [0.01]** | [0.00]*** | -- | [0.22] | [0.35] | [0.35] | [0.35] | -- | -- |
| $R^2$ | 0.038 | 0.052 | 0.062 | 0.067 | 0.101 | 0.152 | 0.127 | 0.151 | 0.136 | 0.136 | 0.175 | 0.175 |
| Observations | 5,059 | 5,059 | 5,059 | 5,059 | 5,059 | 5,059 | 5,059 | 5,059 | 5,059 | 5,059 | 5,059 | 5,059 |
| *Panel C: 2SLS* | | | | | | | | | | | | |
| Years of Schooling | 0.021 | 0.032** | 0.001 | 0.013 | 0.231*** | -0.016 | 0.032* | 0.023** | 0.054** | 0.037** | 0.103*** | 0.040* |
| | (0.019) | (0.013) | (0.009) | (0.011) | (0.041) | (0.043) | (0.017) | (0.012) | (0.022) | (0.015) | (0.036) | (0.021) |
| W-Y p-value | [0.78] | [0.25] | [0.91] | [0.78] | [0.00]*** | -- | [0.08]* | [0.08]* | [0.07]* | [0.07]* | -- | -- |
| 1st Stage F-stat | 81.831 | 81.831 | 81.831 | 81.831 | 81.831 | 81.831 | 81.831 | 81.831 | 81.831 | 81.831 | 81.831 | 81.831 |
| $R^2$ | 0.009 | -- | 0.062 | 0.063 | -- | 0.140 | 0.075 | 0.117 | -- | 0.066 | 0.052 | 0.133 |
| Observations | 5,059 | 5,059 | 5,059 | 5,059 | 5,059 | 5,059 | 5,059 | 5,059 | 5,059 | 5,059 | 5,059 | 5,059 |

*Notes:* Data source is Health and Aging in Africa: A Longitudinal Study. Panel A reports Probit estimates. Panel B presents OLS estimates and Panel C presents 2SLS estimates using log household school expenditures as the instrument. Each estimation contains in its specification: indicators for gender, age, indicators for marital status, indicators for whether the individual's father attended any type of school, indicators for country of origin, dummies for the villages the respondents reside in (they are encoded anonymously for a total of 27 villages). (a) Variables are binary and are coded as such: 1 if the social interaction happens at least once per week and 0 otherwise. Note that tech refers to utilizing technology such as phone, by SMS, through email or the internet. (b) Communication spending includes postage, internet, telephone, mobile phone and others in the past 30 days. (c) Variables are binary and are coded as such: 1 if the respondent strongly agrees with the social perception, hence displaying social cohesion and 0 otherwise. (d) Variables are binary and are coded as such: 1 if respondent partakes in the activity 0 otherwise. Each aggregated mechanism index is the first principal component of all items related to the categorization after utilizing a principal components analysis. The reported 1st Stage F-stat is the Kleiberger-Paap Wald F-statistic. Standard errors are clustered at the village level. W-Y p-values in square brackets represent Westfall-Young adjusted p-values computed to mitigate the issue of multiple hypotheses testing.

*** Significant at the 1 percent level. ** Significant at the 5 percent level. * Significant at the 10 percent level.

**Table 6:** Mechanisms of the Effects of Schooling on Possible Mediating Variables (Health, Entertainment and Stress)

| | Health Habits | | | | | Entertainment | | Stress Proxies | | | | |
|---|---|---|---|---|---|---|---|---|---|---|---|---|
| | Smokes (=1 if Yes)[a] | Alcohol (=1 if Yes)[a] | Log Spending Healthcare[b] | Log Spending Fruits and Vegetables[c] | Health Habits Index | Log Spending Ceremonies[d] | Log Spending Entertainment[e] | Worry[f] | Sadness[f] | Stress[f] | Anger[f] | Stress Index |
| | (2) | (3) | (4) | (5) | (6) | (7) | (8) | (9) | (10) | (11) | (12) | (13) |
| *Panel A: Probit* | | | | | | | | | | | | |
| Years of Schooling | -0.032*** (0.008) | -0.031*** (0.006) | -- | -- | -- | -- | -- | -0.012* (0.006) | -0.006 (0.007) | -0.012 (0.007) | -0.010 (0.009) | -- |
| W-Y p-value | [0.00]*** | [0.00]*** | -- | -- | -- | -- | -- | [0.63] | [0.69] | [0.46] | [0.66] | -- |
| Observations | 5,059 | 5,059 | -- | -- | -- | -- | -- | 5,059 | 5,059 | 5,059 | 5,059 | -- |
| *Panel B: OLS* | | | | | | | | | | | | |
| Years of Schooling | 0.005*** (0.001) | -0.008*** (0.002) | 0.051*** (0.011) | 0.039*** (0.003) | 0.022*** (0.003) | 0.071*** (0.008) | 0.002 (0.002) | -0.002 (0.001) | -0.001 (0.001) | -0.002* (0.001) | -0.001 (0.001) | -0.009** (0.003) |
| W-Y p-value | [0.00]*** | [0.00]*** | [0.00]*** | [0.00]*** | -- | [0.00]*** | [0.26] | [0.67] | [0.71] | [0.56] | [0.67] | -- |
| $R^2$ | 0.155 | 0.149 | 0.068 | 0.090 | 0.196 | 0.068 | 0.021 | 0.018 | 0.009 | 0.015 | 0.010 | 0.025 |
| Observations | 5,059 | 5,059 | 5,059 | 5,059 | 5,059 | 5,059 | 5,059 | 5,059 | 5,059 | 5,059 | 5,059 | 5,059 |
| *Panel C: 2SLS* | | | | | | | | | | | | |
| Years of Schooling | -0.011 (0.007) | -0.019 (0.014) | 0.736*** (0.125) | 0.017 (0.046) | 0.054** (0.027) | 0.765*** (0.136) | 0.093*** (0.033) | -0.024** (0.010) | -0.021** (0.009) | -0.022*** (0.008) | -0.025*** (0.006) | -0.144*** (0.035) |
| W-Y p-value | [0.33] | [0.33] | [0.00]*** | [0.72] | -- | [0.00]*** | [0.05]* | [0.18] | [0.18] | [0.13] | [0.00]*** | -- |
| 1st Stage F-stat | 81.831 | 81.831 | 81.831 | 81.831 | 81.831 | 81.831 | 81.831 | 81.831 | 81.831 | 81.831 | 81.831 | 81.831 |
| $R^2$ | 0.148 | 0.140 | -- | 0.083 | 0.182 | -- | -- | -- | -- | -- | -- | -- |
| Observations | 5,059 | 5,059 | 5,059 | 5,059 | 5,059 | 5,059 | 5,059 | 5,059 | 5,059 | 5,059 | 5,059 | 5,059 |

*Notes:* Data source is Health and Aging in Africa: A Longitudinal Study. Panel A reports Probit estimates. Panel B presents OLS estimates and Panel C presents 2SLS estimates using log household school expenditures as the instrument. Each estimation contains in its specification: indicators for gender, age, indicators for marital status, indicators for whether the individual's father attended any type of school, indicators for country of origin, dummies for the villages the respondents reside in (they are encoded anonymously for a total of 27 villages). (a) Variables are binary and are coded as such: 1 if respondent partakes in the activity 0 otherwise. (b) Healthcare spending includes doctor, nurse, hospital, medicines etc. in the past 12 months. (c) Money spent on fruits and vegetables includes apple, banana, guava, avocado, orange, naartjie, lemon, grape, raisin, melon, grapefruit, pineapple, mulberry, wild fruit, butternut, carrot, spinach, muroho, cabbage, broccoli etc. in the past 30 days (d) Ceremony spending includes spending on festivals and religious ceremonies such as festival clothes gifts, foods, etc. in the past 12 months. (e) Entertainment spending includes spending on games, table pool stadium soccer matches and attending bashes in the past 30 days. (f) Variables are binary and coded as such: 1 if respondent experiences the particular emotion the day before the interview, and 0 if not. Each aggregated mechanism index is the first principal component of all items related to the categorization after utilizing a principal components analysis. The reported 1st Stage F-stat is the Kleiberger-Paap Wald F-statistic. Standard errors are clustered at the village level. W-Y p-values in square brackets represent Westfall-Young adjusted p-values computed to mitigate the issue of multiple hypotheses testing.

*** Significant at the 1 percent level. ** Significant at the 5 percent level. * Significant at the 10 percent level.

**Table 7:** Robustness Analysis: Effects of Schooling on Cognitive Performance (HAALSI)

|  | Immediate Episodic Memory | Delayed Episodic Memory | Orientation | Aggregate Memory | Executive Functioning | Attention and Speed | Aggregate Cognition |
|---|---|---|---|---|---|---|---|
|  | (1) | (2) | (3) | (4) | (5) | (6) | (7) |
| *Panel A: 2SLS Number of Siblings* | | | | | | | |
| Years of Schooling | 0.292** | 0.213** | 0.211*** | 0.290** | 0.140** | 0.179*** | 0.299*** |
|  | (0.124) | (0.098) | (0.075) | (0.121) | (0.066) | (0.063) | (0.103) |
| 1st Stage F-stat | 12.897 | 12.897 | 12.897 | 12.897 | 12.897 | 12.897 | 12.897 |
| $R^2$ | -- | -- | -- | -- | 0.057 | -- | -- |
| Observations | 5,059 | 5,059 | 5,059 | 5,059 | 5,059 | 5,059 | 5,059 |
| *Panel B: 2SLS Money Spent on Education Indicator* | | | | | | | |
| Years of Schooling | 0.256*** | 0.330*** | 0.222*** | 0.318*** | 0.066 | 0.258*** | 0.323*** |
|  | (0.057) | (0.095) | (0.044) | (0.084) | (0.054) | (0.049) | (0.054) |
| 1st Stage F-stat | 39.414 | 39.414 | 39.414 | 39.414 | 39.414 | 39.414 | 39.414 |
| $R^2$ | -- | -- | -- | -- | 0.134 | -- | -- |
| Observations | 5,059 | 5,059 | 5,059 | 5,059 | 5,059 | 5,059 | 5,059 |

*Notes:* Data source is Health and Aging in Africa: A Longitudinal Study. Each coefficient and standard error in each cell represents a different regression. Panel A reports 2SLS estimates of years of education on cognition performance using the number of siblings as the instrument; Panel B reports 2SLS estimates using money spent on education indicator as the instrument. The reported 1st Stage F-stat is the Kleiberger-Paap Wald F-statistic. Aggregate Memory is the first principal component of all memory-related test scores. Aggregate Cognition is the first principal component of all cognition test scores. Each cognitive test measure is age-adjusted. Each regression contains in its specification: indicators for gender, age, indicators for marital status, indicators for whether the individual's father attended any type of school, indicators for country of origin, dummies for the villages the respondents reside in (they are encoded anonymously for a total of 27 villages). Standard errors are clustered at the village level.

*** Significant at the 1 percent level. ** Significant at the 5 percent level. * Significant at the 10 percent level.

# Appendix A

**Table A.1:** 2SLS Effects of Schooling on Performance in Various Cognitive Domains by Sample (HAALSI)

|  | Immediate Episodic Memory | Delayed Episodic Memory | Orientation | Aggregate Memory Index | Executive Functioning | Attention and Speed | Aggregate Cognition Index |
|---|---|---|---|---|---|---|---|
|  | (1) | (2) | (3) | (4) | (5) | (6) | (7) |
| *Panel A: S.A.-born Sample* | | | | | | | |
| Years of Schooling | 0.165*** (0.032) | 0.235*** (0.061) | 0.098*** (0.026) | 0.222*** (0.048) | 0.092*** (0.031) | 0.091*** (0.034) | 0.191*** (0.036) |
| 1st Stage F-stat | 91.015 | 91.015 | 91.015 | 91.015 | 91.015 | 91.015 | 91.015 |
| $R^2$ | -- | -- | 0.095 | -- | 0.097 | 0.040 | -- |
| Observations | 3531 | 3531 | 3531 | 3531 | 3531 | 3531 | 3531 |
| *Panel B: Foreign-born Sample* | | | | | | | |
| Years of Schooling | 0.387*** (0.132) | 0.417** (0.168) | 0.518*** (0.170) | 0.436*** (0.139) | 0.205 (0.167) | 0.612** (0.270) | 0.600*** (0.172) |
| 1st Stage F-stat | 8.519 | 8.519 | 8.519 | 8.519 | 8.519 | 8.519 | 8.519 |
| $R^2$ | -- | -- | -- | -- | 0.000 | -- | -- |
| Observations | 1528 | 1528 | 1528 | 1528 | 1528 | 1528 | 1528 |

*Notes:* Data source is Health and Aging in Africa: A Longitudinal Study. Each coefficient and standard error in each cell represents a different regression. Panel A reports 2SLS estimates using log household school expenditures as the instrument for the S.A. sample; Panel B reports 2SLS estimates using log household school expenditures as the instrument for the Non-S.A. sample. The reported 1st Stage F-stat is the Kleiberger-Paap Wald F-statistic. Aggregate Memory is the first principal component of all memory-related test scores. Aggregate Cognition is the first principal component of all cognition test scores. Each cognitive test measure is age-adjusted. Each regression contains in its specification: indicators for gender, age, indicators for marital status, indicators for whether the individual's father attended any school, indicators for country of origin, dummies for the villages the respondents reside in (they are encoded anonymously for a total of 27 villages). Standard errors are clustered at the village level.

*** Significant at the 1 percent level. ** Significant at the 5 percent level. * Significant at the 10 percent level.

**Table A.2:** Effects of Schooling on Performance in Various Cognitive Domains with No Outliers (HAALSI)

|  | Immediate Episodic Memory | Delayed Episodic Memory | Orientation | Aggregate Memory Index | Executive Functioning | Attention and Speed | Aggregate Cognition Index |
|---|---|---|---|---|---|---|---|
|  | (1) | (2) | (3) | (4) | (5) | (6) | (7) |
| *Panel A: OLS* | | | | | | | |
| Years of Schooling | 0.057*** | 0.060*** | 0.064*** | 0.065*** | 0.066*** | 0.048*** | 0.080*** |
|  | (0.006) | (0.005) | (0.006) | (0.006) | (0.005) | (0.004) | (0.006) |
| $R^2$ | 0.157 | 0.105 | 0.179 | 0.155 | 0.132 | 0.193 | 0.243 |
| Observations | 4997 | 4997 | 4997 | 4997 | 4997 | 4997 | 4997 |
| *Panel B: 2SLS* | | | | | | | |
| Years of Schooling | 0.192*** | 0.245*** | 0.160*** | 0.241*** | 0.110*** | 0.166*** | 0.244*** |
|  | (0.028) | (0.055) | (0.031) | (0.043) | (0.032) | (0.036) | (0.031) |
| 1st Stage F-stat | 62.721 | 62.721 | 62.721 | 62.721 | 62.721 | 62.721 | 62.721 |
| $R^2$ | -- | -- | 0.066 | -- | 0.108 | 0.018 | -- |
| Observations | 4997 | 4997 | 4997 | 4997 | 4997 | 4997 | 4997 |

*Notes:* Data source is Health and Aging in Africa: A Longitudinal Study. Each coefficient and standard error in each cell represents a different regression. Years of Schooling and Log Schooling Fee outliers are removed from this sample. Panel A reports OLS estimates of years of education on cognitive performance; Panel B reports 2SLS estimates using log household school expenditures as the instrument. The reported 1st Stage F-stat is the Kleiberger-Paap Wald F-statistic. Aggregate Memory is the first principal component of all memory-related test scores. Aggregate Cognition is the first principal component of all cognition test scores. Each cognitive test measure is age-adjusted. Each regression contains in its specification: indicators for gender, age, indicators for marital status, indicators for whether the individual's father attended any school, indicators for country of origin, dummies for the villages the respondents reside in (they are encoded anonymously for a total of 27 villages). Standard errors are clustered at the village level.

*** Significant at the 1 percent level. ** Significant at the 5 percent level. * Significant at the 10 percent level.